\documentclass[sigconf]{acmart}
\usepackage{url}
\makeatletter
\let\orig@maketitle\maketitle           % save original
\renewcommand{\maketitle}{%             % new wrapper
  \begingroup
    \let\OldUrlFont\UrlFont             % 1) remember global setting
    \renewcommand\UrlFont{\ttfamily\small}% 2) narrow just for banner
    \orig@maketitle                     % 3) run real maketitle
  \endgroup
  \let\UrlFont\OldUrlFont               % 4) restore for rest of paper
}
\makeatother

\renewcommand\footnotetextcopyrightpermission[1]{} % removes footnote with conference information in first column
\AtBeginDocument{%
  }

\usepackage{enumitem}

\begin{document}

%%
%% The "title" command has an optional parameter,
%% allowing the author to define a "short title" to be used in page headers.
\title{LayLens: Improving Deepfake Understanding through Simplified Explanations}

%%
%% The "author" command and its associated commands are used to define
%% the authors and their affiliations.
%% Of note is the shared affiliation of the first two authors, and the
%% "authornote" and "authornotemark" commands
%% used to denote shared contribution to the research.

\author{Abhijeet Narang}
\email{anar0033@student.monash.edu}
\orcid{0000-0003-0677-910X}
\affiliation{%
  \institution{Monash University}
  \city{Melbourne}
  \country{Australia}
}

\author{Parul Gupta}
\email{parul@monash.edu}
\orcid{0000-0002-4379-1573}
\affiliation{%
  \institution{Monash University}
  \city{Melbourne}
  \country{Australia}
}

\author{Liuyijia Su}
\email{lsuu0008@student.monash.edu}
\orcid{0009-0002-4929-2326}
\affiliation{%
  \institution{Monash University}
  \city{Melbourne}
  \country{Australia}
}

\author{Abhinav Dhall}
\email{abhinav.dhall@monash.edu}
\orcid{0000-0002-2230-1440}
\affiliation{%
  \institution{Monash University}
  \city{Melbourne}
  \country{Australia}
}

% \authorsaddresses{%
%   Abhijeet Narang, anar0033@student.monash.edu;%
%   Parul Gupta, parul@monash.edu;%
%   Liuyijia Su, lsuu0008@student.monash.edu;%
%   Abhinav Dhall, abhinav.dhall@monash.edu%
% }

%%
%% By default, the full list of authors will be used in the page
%% headers. Often, this list is too long, and will overlap
%% other information printed in the page headers. This command allows
%% the author to define a more concise list
%% of authors' names for this purpose.
\renewcommand{\shortauthors}{Narang et al.}

%%
%% The abstract is a short summary of the work to be presented in the
%% article.
% \begin{teaserfigure}
%   \centering
%   \includegraphics[width=\linewidth]{samples/images/teaser.pdf}
%   \caption{Comparison View: The user-uploaded fake image is analyzed by the system to localize regions suspected of manipulation (left). Based on these findings, a plausible reconstruction of the original, unaltered image is generated (right). This side-by-side view facilitates intuitive visual understanding of the manipulated content.}
%   \Description{.}
%   \label{fig:comparison}
% \end{teaserfigure}
\begin{teaserfigure}
    \centering
    \includegraphics[width=\linewidth]{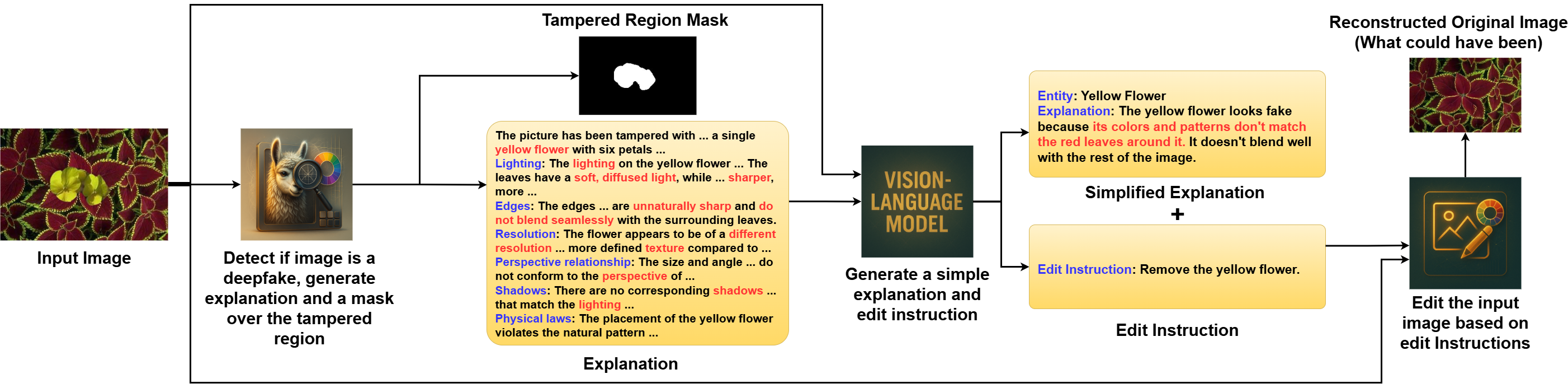}
    % \vspace{-1em}
    \caption{LayLens makes identifying and understanding deepfakes more accessible and easier to understand, by (a) transforming the long and complex explanations into simple, non-technical reasoning, and (b) re-imagining the fake image by removing the manipulated region, resulting in a version of what the original image may have looked like.}
    \label{fig:flowDiagram}
    % \vspace{-1em}
    \Description{.}
\end{teaserfigure}
\begin{abstract}
This demonstration paper presents \textbf{LayLens}, a tool aimed to make deepfake understanding easier for users of all educational backgrounds. While prior works often rely on outputs containing technical jargon, LayLens bridges the gap between model reasoning and human understanding through a three-stage pipeline: (1) explainable deepfake detection using a state-of-the-art forgery localization model, (2) natural language simplification of technical explanations using a vision-language model, and (3) visual reconstruction of a plausible original image via guided image editing. The interface presents both technical and layperson-friendly explanations in addition to a side-by-side comparison of the uploaded and reconstructed images. A user study with 15 participants shows that simplified explanations significantly improve clarity and reduce cognitive load, with most users expressing increased confidence in identifying deepfakes. LayLens offers a step toward transparent, trustworthy, and user-centric deepfake forensics.
\end{abstract}
% \vspace{-5pt}

%%
%% The code below is generated by the tool at http://dl.acm.org/ccs.cfm.
%% Please copy and paste the code instead of the example below.
%%


%%
%% Keywords. The author(s) should pick words that accurately describe
%% the work being presented. Separate the keywords with commas.
%% A "teaser" image appears between the author and affiliation
%% information and the body of the document, and typically spans the
%% page.
% \begin{teaserfigure}
%   \includegraphics[width=\textwidth]{sampleteaser}
%   \caption{Seattle Mariners at Spring Training, 2010.}
%   \Description{Enjoying the baseball game from the third-base
%   seats. Ichiro Suzuki preparing to bat.}
%   \label{fig:teaser}
% \end{teaserfigure}

% \received{7 July 2025}
% \received[revised]{12 March 2009}
% \received[accepted]{5 June 2009}

%%
%% This command processes the author and affiliation and title
%% information and builds the first part of the formatted document.
\maketitle
% \vspace*{-0.7em\baselineskip}

\section{Introduction}
The advancement in generative AI technology has led to a proliferation of AI-generated and manipulated content (commonly referred to as \textit{deepfakes}), raising significant challenges for media authenticity, public trust and digital safety. While some of the state-of-the-art deepfake detection tools provide textual explanations of why an image may be a fake, their outputs are often opaque and overly technical for non-expert users. To address this critical gap, we present \textbf{LayLens}: an intuitive, web-based interface that allows users to upload an image and receive highly simplified, visually guided explanations of why the image may be fake, along with a plausible reconstruction of what the authentic image might have looked like. Our goal is to bridge the gap between high-performance deepfake detection and public interpretability by offering explanations that are not only accurate but also immediately understandable to general audiences, including educators, journalists, content moderators and everyday users.

\section{Related Work}
% Several prior approaches have addressed the detection and explanation of deepfake content.
Traditional deepfake detection models, such as ~\cite{zhou2018learning, artifact_2019, chugh_2020} primarily focused on obtaining high accuracy in detecting manipulated inputs but lacked intuitive interpretability for lay users.
In earlier years, explainability in deepfake detection was introduced through saliency-based techniques like LRP~\cite{LRP_2015}, Grad-CAM~\cite{grad_cam_2017}, LIME~\cite{lime} and SHAP~\cite{shap_2017}, which produce saliency maps highlighting image regions most influential to the classifier’s output. These approaches are model-agnostic and were adopted to identify the important input regions in deepfake detection models in works such as ~\cite{DBLP:journals/corr/abs-2105-05902, 9092227}. Later works such as ~\cite{Soltandoost_2025_WACV} propose identifying embeddings corresponding to local facial regions in images, and use them to produce interpretable classification outputs. Recently, Vision-Language model based approaches, such as FakeShield ~\cite{xu2024fakeshield} and SIDA ~\cite{SIDA_2025_CVPR} have become popular, which provide textual rationale along with localization of the tampered region. Despite these advancements, current explainable detection tools exhibit gaps that limit their usefulness for non-experts. A common issue is that the explanations are too low-level or technical. There is an evident gap in the literature for a deepfake detection interface that provides simplified yet informative visual and textual explanations accessible to the general public.

\section{Interface Design}
% \section{Flow Diagram}
% \begin{figure*}[h]
%     \centering
%     \includegraphics[width=\linewidth]{samples/images/Laylens.png}
%     \vspace{-2em}
%     \caption{Flow Diagram for LayLens.}
%     \label{fig:flowDiagram}
%     \vspace{-1em}
%     \Description{.}
% \end{figure*}
\begin{figure}
  \centering
  \includegraphics[width=\linewidth]{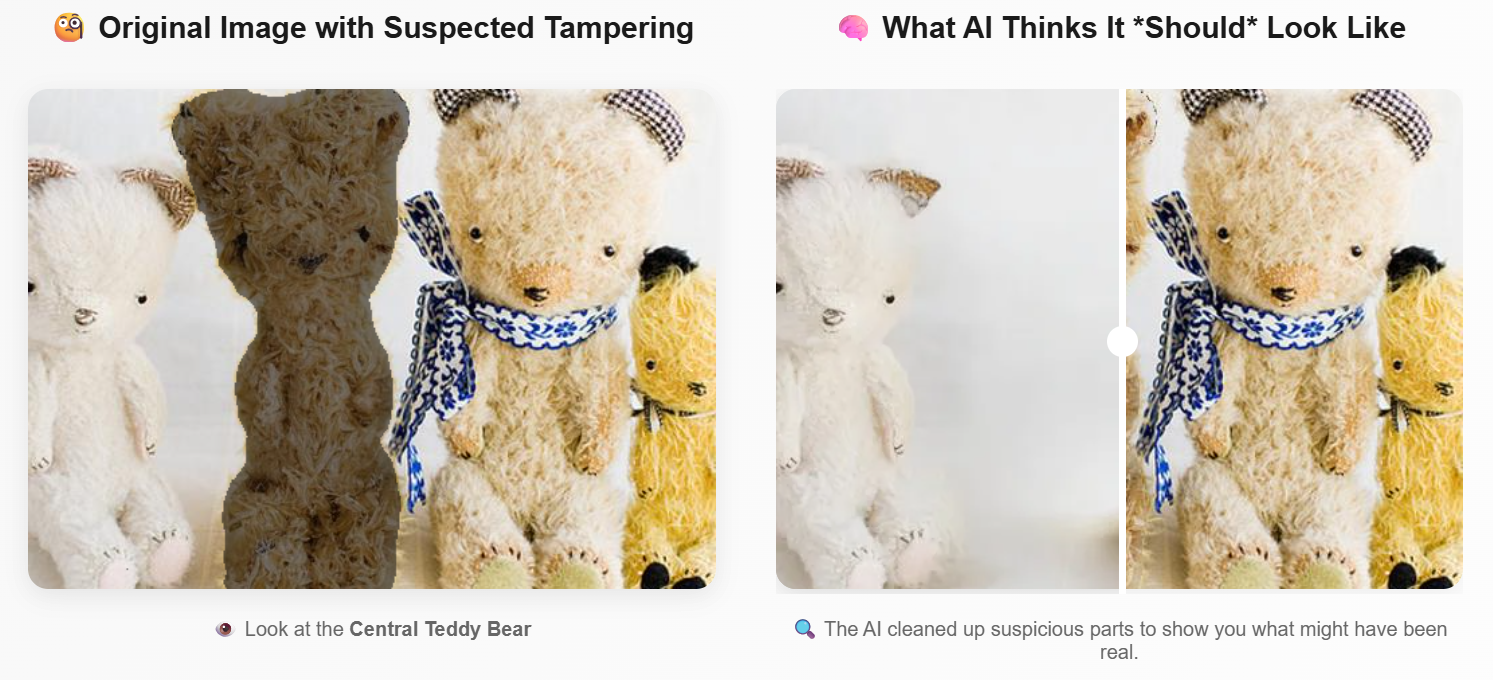}
  \caption{Comparison View: The user-uploaded fake image is analyzed by the system to localize regions suspected of manipulation (left). Based on these findings, a plausible reconstruction of the original, unaltered image is generated (right). This side-by-side view facilitates intuitive visual understanding of the manipulated content.}
  \Description{.}
  \label{fig:comparison}
\end{figure}
\begin{figure}
  \centering
  \includegraphics[width=\linewidth]{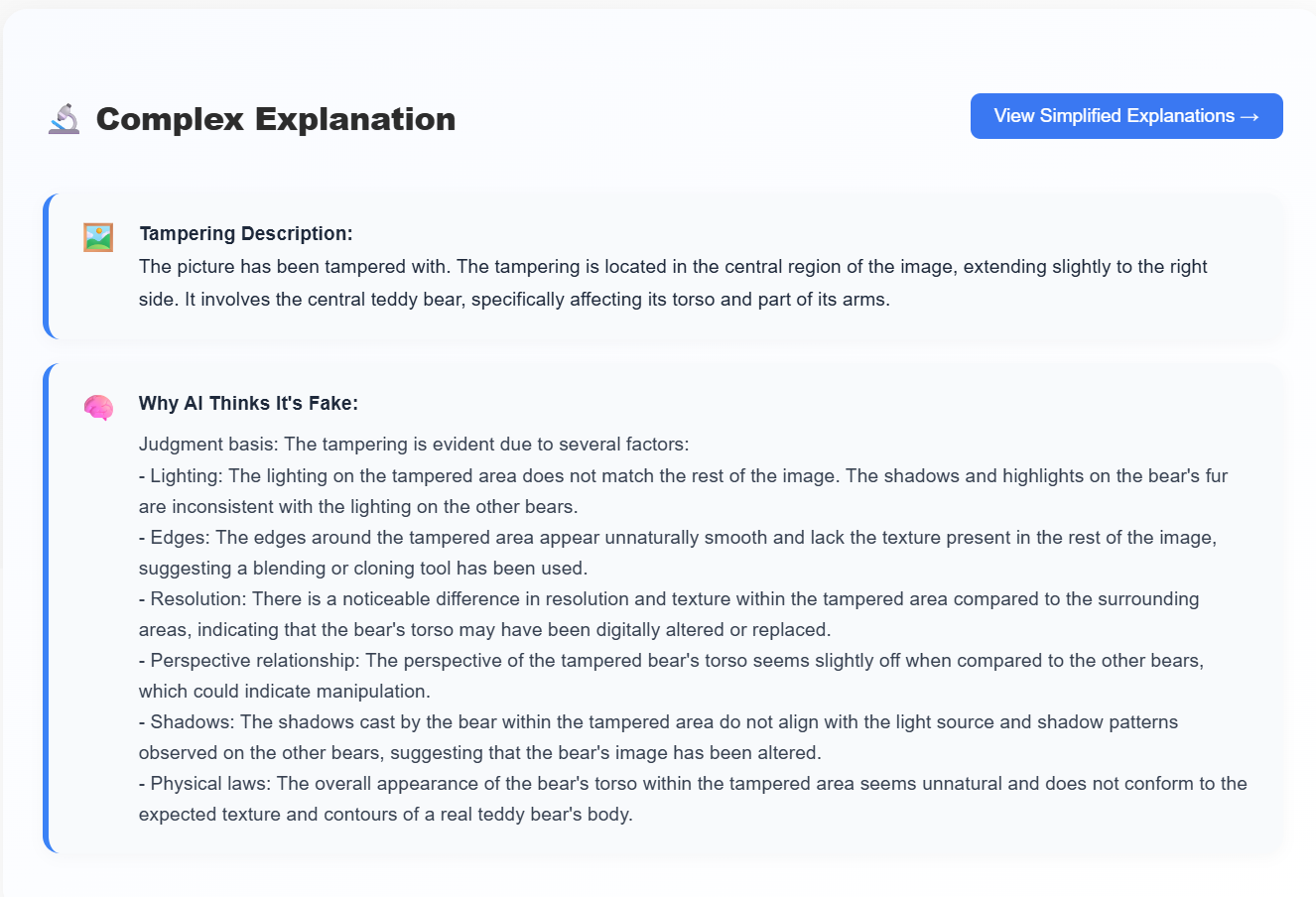}
  \caption{Complex Explanation: This view provides a detailed breakdown of several technical factors such as lighting inconsistencies, resolution artifacts, perspective anomalies, and shadow discrepancies that inform the model's decision in identifying potential manipulations within the image.}
  \Description{.}
  \label{fig:complex}
\end{figure}
\begin{figure}
  \centering
  \includegraphics[width=\linewidth]{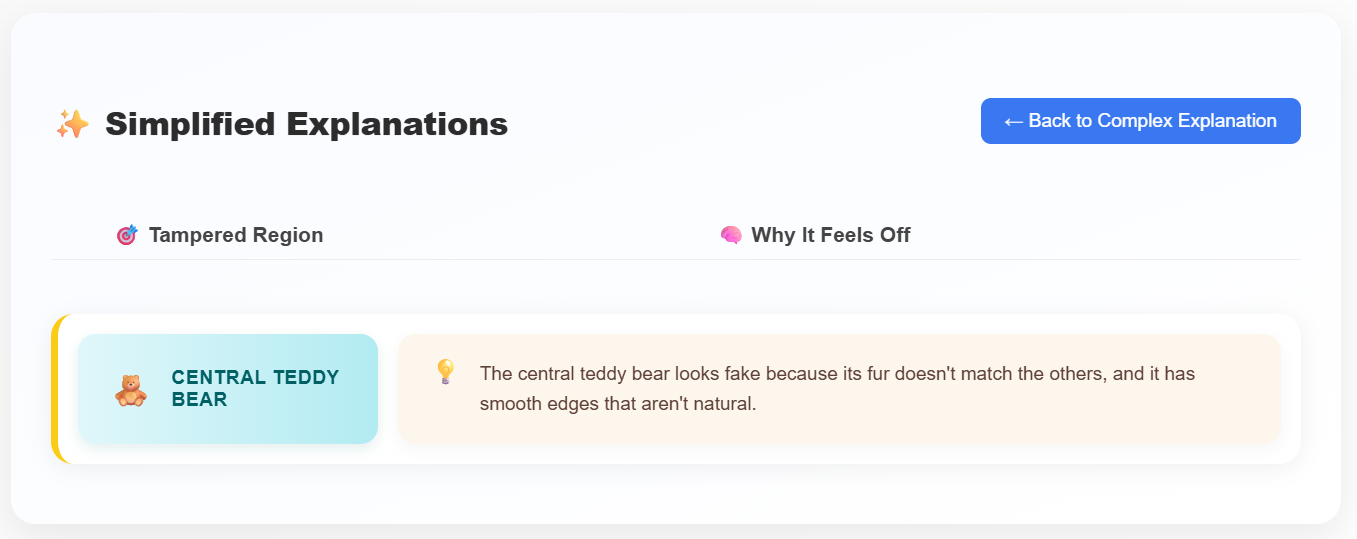}
  \caption{Simplified Explanation: This view translates the technical explanation into concise, region-level descriptions tailored for non-expert users. By reducing cognitive load and using accessible language, it enhances interpretability and user engagement.}
  \Description{.}
  \label{fig:simplified}
\end{figure}

\begin{figure*}[h]
    \centering
    \includegraphics[width=0.98\linewidth]{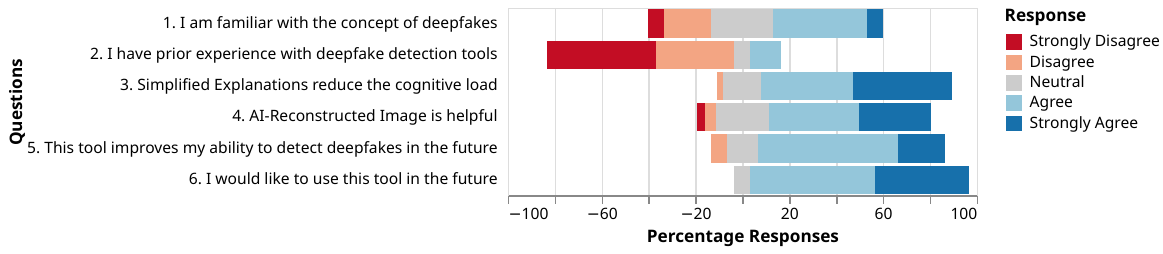}
    \vspace{-1em}
    \caption{Distribution of User survey likert scale responses for various questions.}
    \label{fig:likert_visualize}
    % \vspace{-1em}
    \Description{.}
\end{figure*}

\textbf{LayLens} is designed as an interactive, end-to-end system that not only detects deepfake images but also presents explanations in a form accessible to both technical and non-technical users. The system integrates state-of-the-art methods in deepfake detection, localization and explanation, along with generative image editing, and wraps them in a user-friendly interface that encourages interpretability and engagement. (Figure~\ref{fig:flowDiagram}) gives an overview of this three-stage pipeline, showing how LayLens moves from deepfake detection, through explanation simplification, to visual reconstruction. After the user uploads an image to the interface, the following components are triggered:

\subsubsection{Comparison View (Figure~\ref{fig:comparison}):} First, the users are presented with a side-by-side visual display. The left panel shows the original uploaded image, overlaid with a softly pulsating mask which highlights regions suspected to have been manipulated. The right panel shows a reconstructed version of the image, generated based on the AI's understanding of what the non-tampered image could have looked like. A slider enables intuitive, pixel-level comparison between the two. While the mask is obtained through Fakeshield~\cite{xu2024fakeshield}, Step1X-Edit~\cite{liu2025step1x-edit} creates the imagined version of the original image, by taking the user-uploaded image along with the edit instruction (as obtained in Section \ref{sec:explanation}(\ref{item:simplified}) below) as inputs. We also tried using ICEdit~\cite{zhang2025ICEdit} here, but found Step1X-Edit's editing performance to be better (based upon manual observation).

\subsubsection{Explanation Card:}\label{sec:explanation} Below the comparison view, users can access a flip-style explanation card. This card presents the system's reasoning in two tiers:

\begin{enumerate}[wide, labelwidth=!, labelindent=0pt]
    \item\textbf{Complex Explanation (Figure~\ref{fig:complex}):} Offers detailed reasoning behind the detection decision, including references to lighting inconsistencies, perspective errors, shadow artifacts, resolution discrepancies and physically implausible structures. This is obtained from the output of the Fakeshield~\cite{xu2024fakeshield} model. Here, we also experimented with SIDA~\cite{SIDA_2025_CVPR}, but found Fakeshield's explanations to be more intuitive and accurate through manual observation.
    \item \label{item:simplified}\textbf{Simplified Explanation (Figure~\ref{fig:simplified}):} Performs automatic text simplification on complex explanation. Each identified region (e.g., \textit{``Central Teddy Bear''}) is paired with a relevant emoji and a simplified explanation, generated by a VLM~\cite{bai2025qwen25vltechnicalreport}. It is prompted with both the image and FakeShield's technical explanation and tasked with generating a structured, human-readable JSON output. For each manipulated region, the VLM outputs:
    \begin{itemize}
        \item A simplified explanation of why the region appears fake,
        \item An associated emoji to visually cue the user,
        \item A concise edit instruction describing how to correct or restore the manipulated region.
    \end{itemize}

\end{enumerate}

\noindent Thus, LayLens offers a complete workflow that lets the users choose their preferred level of technical detail while always grounding its decisions in visual evidence. It enhances the accessibility, transparency, and interpretability of deepfake detection for a broad spectrum of users, from forensic analysts to everyday citizens. This design lets different users choose their preferred explanation detail level, while always seeing a visual demonstration of the AI’s reasoning.

\section{User Survey}
To evaluate the effectiveness of LayLens in providing accessible explanations for deepfake images, we conducted a user study with 15 participants, of whom 11 were familiar with the concept of deepfakes, and 3 had prior experience with deepfake detection tools. Each participant interacted with the system by analyzing 10 AI-manipulated images. Overall, users preferred the simplified explanations over the complex ones in 65.3\% of the cases. Notably, in 81.3\% of comparisons, participants reported that the simplified explanations reduced their cognitive load in understanding why an image might be a deepfake. The side-by-side visualization, featuring the uploaded (potentially fake) image alongside a plausibly reconstructed original, was considered helpful in enhancing understanding in 69.3\% of the instances. Furthermore, 80\% of participants indicated that the experience improved their confidence in detecting deepfakes in the future, and 93.3\% expressed interest in using such a tool for identifying manipulated media going forward. The distribution of responses across all questions, measured using a 5-point Likert scale, is illustrated in Figure~\ref{fig:likert_visualize}. Additionally, we performed a Wilcoxon signed-rank test to assess perceived changes in \textbf{Ease of Understanding}, \textbf{Clarity} and \textbf{Accuracy} when switching from complex to simplified explanations. The resulting p-values were $3.25e-06$, $0.01$ and $0.30$, respectively. These results indicate statistically significant improvements (p $\le 0.05$) in both \textbf{Ease of Understanding} and \textbf{Clarity} when simplified explanations are presented.

\section{Conclusion}
In this work, we presented LayLens, a system to make deepfake detection more accessible, interpretable and engaging for users with varying levels of technical expertise. By integrating detection, natural language simplification, and generative reconstruction, LayLens enables both technical and non-expert users to understand why an image may be fake and what the original might have looked like. Our user study shows that simplified, visually grounded explanations reduce cognitive load and enhance user confidence. LayLens demonstrates that deepfake detection can be made both accurate and accessible, paving the way for more transparent and trustworthy media forensics. As a future work, we will develop an open-source, efficient and explainable model for large-scale audio-visual deepfakes~\cite{cai2025avdeepfake1mlargescaleaudiovisualdeepfake} and further, extend it to incorporate multi-lingual, code-switched videos~\cite{kuckreja2025tellhabibirealfake}.
\bibliographystyle{ACM-Reference-Format}
\bibliography{sample-sigconf}

%%
%% If your work has an appendix, this is the place to put it.
% \appendix

% \section{Research Methods}

\end{document}